\documentclass[english,aps,floats,twocolumn,showpacs, nofootinbib,floatfix]{revtex4}
\usepackage{pslatex}
\usepackage[T1]{fontenc}
\usepackage[latin1]{inputenc}
\usepackage{graphicx}
\usepackage{epsfig}
\usepackage{longtable}
\usepackage{float}

{
{
{
\newcommand{\bea}{\begin{eqnarray}}
\newcommand{\eea}{\end{eqnarray}}

\newcommand{\nc}{\newcommand}
\nc{\renc}{\renewcommand}
\nc{\eqs}[2]{\mbox{Eqs.~(\ref{#1},\,\ref{#2})}}
\nc{\eq}[1]{\mbox{Eq.~(\ref{#1})}}
\nc{\figs}[2]{\mbox{Figs.~(\ref{#1},\,\ref{#2})}}
\nc{\fig}[1]{\mbox{Fig~.(\ref{#1})}}
\nc{\be}[1]{\begin{equation} \mbox{$\label{#1}$}}
\nc{\ee}{\vspace{0.1cm}\end{equation}}

\newcommand{\bean}{\begin{eqnarray*}}
\newcommand{\eean}{\end{eqnarray*}}

\def\bfn{{\bf n}}
\def\bfp{{\bf p}}

\def\lae{\;^{<}_{\sim} \;}

\def\cP{{\cal P}}

\begin{document}
\title{Negative Running of the Spectral Index, Hemispherical Asymmetry and the Consistency of Planck with Large $r$}
\author{John McDonald}
\email{j.mcdonald@lancaster.ac.uk}
\affiliation{Dept. of Physics, University of 
Lancaster, Lancaster LA1 4YB, UK}

\begin{abstract}

     Planck favours a negative running of the spectral index,  with the likelihood being dominated by low multipoles $l \lae 50$ and no preference for running at higher $l$. A negative spectral index is also necessary for the 2-$\sigma$ Planck upper bound on the tensor-to-scalar ratio $r$ to be consistent with values significantly larger than 0.1. Planck has also observed a hemispherical asymmetry of the CMB power spectrum, again mostly at low multipoles. Here we consider whether the physics responsible for the hemispherical asymmetry could also account for the negative running of the spectral index and the consistency of Planck with a large value of $r$. A negative running of the spectral index can be generated if the hemispherical asymmetry is due to a scale- and space-dependent modulation which suppresses the CMB power spectrum at low multipoles. We show that the observed hemispherical asymmetry at low $l$ can be generated while satisfying constraints on the asymmetry at higher $l$ and generating a negative spectral index of the right magnitude to account for the Planck observation and to allow Planck to be consistent with a large value of $r$.

\end{abstract}
 \maketitle

\section{Introduction}

    Planck has observed both a hemispherical asymmetry of the CMB power spectrum \cite{planckasym} and a preference for a negative running of the spectral index \cite{cosmoparams}. Both features are difficult to achieve in conventional inflation models, which typically predict statistical isotropy and a negligibly small value for the running of the spectral index.

   The negative running of the spectral index observed by Planck\footnote{WMAP 9-year observations also find a negative running of the spectral index \cite{wmap9}.} is given by $n_{s}^{\prime} = dn_{s}/d\ln(k) = -0.013 \pm 0.009$ for the case without tensor modes and $n_{s}^{\prime} = -0.021 \pm 0.011$ in the presence of tensor modes\footnote{We will quote Planck + WP values for reference; other values are similar.} \cite{cosmoparams}.
The running of the spectral index obtained in \cite{cosmoparams} is based on a maximum likelihood fit which assumes a scale-independent value for the running. However, it was also found that 
the likelihood of the negative value of the running  
is dominated by low multipoles, $l \lae 50$, with no preference for a negative running at higher $l$ \cite{cosmoparams}. This suggests that the negative running of the spectral index is scale-dependent and largest at low multipoles.

   The hemispherical asymmetry \cite{asym1,wmap5,hansennew} is also dominated by low multipoles. The asymmetry can be parameterized by a temperature dipole \cite{gordon} 
\be{e0} \frac{\delta T}{T}(\hat{\bfn}) = \left(\frac{\delta T}{T}\right)_{o}(\hat{\bfn})\left[1 + A\;\hat{\bfn}.\hat{\bfp} \right]   ~,\ee  
where $\left(\frac{\delta T}{T}\right)_{o}(\hat{\bfn})$ is a statistically isotropic temperature fluctuation, $A$ is the magnitude of the asymmetry and $\hat{\bfp}$ is its direction.
The asymmetry is observed by Planck to be $A = 0.073 \pm 0.010$ for multipoles $l = 2-64$ \cite{planckasym}, in agreement with earlier WMAP results \cite{wmap5}. At larger $l$ there is no evidence of an asymmetry \cite{hirata,sh}, with $A < 0.0045$ (95 $\%$ c.l.) at $l = 601-2048$ \cite{sh}.

   The domination of both the hemispherical asymmetry and the negative running of the spectral index by low multipoles  suggests that the same new physics could be responsible for both. This would also explain why a significant negative running of the spectral index is observed when most inflation models predict a negligible running.

       BICEP2 reported the observation of gravity waves from inflation via CMB B-mode polarization at multipoles $l \lae 150$, with tensor-to-scalar ratio $r = 0.16^{+0.06}_{-0.05}$ \cite{bicep1,bicep2}. However, it has since become clear that the dust signal was underestimated in the original analysis
 \cite{bicep1,x1}. A future joint analysis of BICEP2 and Planck data will be necessary in order to establish whether a large tensor-to-scalar ratio exists \cite{x1}. If confirmed, a value of $r$ significantly larger than 0.1 would be in tension with the 95 $\%$ c.l. upper limit from Planck based on the CMB temperature spectrum, $r_{0.002} < 0.11$ \cite{cosmoparams}. However, this upper bound becomes weaker in the presence of a negative running of the spectral index, in which case 
$r_{0.002} < 0.26$ \cite{cosmoparams}. The consistency of Planck and the original BICEP2 result has also been addressed in \cite{comp1,comp2,comp3,comp4,comp5}, while the existence of the tension between them has been critically examined in \cite{fig} and \cite{dust}. A possible link between the tension and the hemispherical asymmetry, via a spatially-varying $r$, has been explored in \cite{kamr}. 
  
  Here we wish to show that a modification of the CMB power spectrum which can account for the hemispherical asymmetry can also generate a negative running of the spectral index of the right magnitude to account for the value observed by Planck. This would also allow for consistency between the Planck upper bound on $r$ and an observed value which is significantly larger than 0.1.

\section{A modulation model for the hemispherical asymmetry and negative running of the spectral index}

   In \cite{agen1} we introduced a power spectrum model for the hemispherical asymmetry based on the addition of a space- and scale-dependent component $\cP_{asy}$ to the adiabatic power spectrum $\cP_{\zeta}$,
\be{e1} \cP_{\zeta} = \cP_{inf} + \cP_{asy} ~,\ee
where $\cP_{inf}$ is the conventional inflaton power spectrum and $\cP_{asy}$ is responsible for the power asymmetry. $\cP_{asy}$ is assumed to consist of a mean value and a spatially-varying dipole term. On the surface of last scattering $\cP_{asy}$ is given by,
\be{e2} \cP_{asy} = \hat{\cP}_{asy} + \Delta \cP_{asy} \; \hat{\bfn}.\hat{\bfp}    ~.\ee 
In \cite{agen1} $\cP_{asy}$ was assumed to be due to a new contribution to the adiabatic perturbation which is uncorrelated with the inflaton perturbation.  In this case $\cP_{asy}$ is strictly positive in sign. This leads to a positive running of the spectral index. To have a negative running, we need $\hat{\cP}_{asy}$ to be negative and so to partially cancel the power from the inflaton perturbations. This is possible if the inflaton power spectrum has a space- and scale-dependent modulation which suppresses its magnitude. For example, if 
\be{e3} \cP_{\zeta} = \frac{\cP_{inf}}{(1 + f_{asy}(\hat{\bfn}))} ~,\ee
where
\be{e4} f_{asy}(\hat{\bfn}) = \hat{f}_{asy} + \Delta f_{asy} \; \hat{\bfn}.\hat{\bfp}   ~,\ee  
then for $f_{asy}(\hat{\bfn}) \ll 1$,   
\be{e4a} \cP_{\zeta} \approx 
\cP_{inf} - f_{asy}(\hat{\bfn}) \cP_{inf} ~.\ee
In this case $\cP_{\zeta}$ has the form of \eq{e1}, with $\hat{\cP}_{asy} = -\hat{f}_{asy} \cP_{inf}$ and 
$\Delta \cP_{asy} = - \Delta f_{asy} \cP_{inf}$.

\subsection{The Hemispherical Asymmetry}

  The observed asymmetry on large angles, which we denote by $A_{large}$, comes from averaging over multipoles $l = 2$ to 64. We will assume that $\hat{\cP}_{asy}$ and $\Delta \cP_{asy}$ have the same scale-dependence, corresponding to a power law dependence on $k$ with spectral index $n_{\sigma}$,
\be{e4b} \hat{\cP}_{asy} = \hat{\cP}_{asy\;0} \left(\frac{k}{k_{0}}\right)^{n_{\sigma}-1} \;\;;\;\; \Delta \hat{\cP}_{asy} = \Delta \hat{\cP}_{asy\;0} \left(\frac{k}{k_{0}}\right)^{n_{\sigma}-1}   ~.\ee
$k$ is approximately related to $l$ by $k = l/x_{ls}$, where $x_{ls} = 14100$ Mpc is the comoving distance to the last-scattering surface. $k_{0}$ is the pivot scale, which we define to be $0.05 \; {\rm Mpc}^{-1}$. This corresponds to $l_{0} \approx 700$. The asymmetry $A$ from modes in the range $l_{min}$ to $l_{max}$ is given by \cite{agen1}
\be{e5} A \approx \frac{|\xi|_{0} (\Delta \cP_{asy}/\hat{\cP}_{asy})_{0}}{2} \times
 \frac{ \displaystyle\sum_{l=l_{min}}^{l_{max}} \frac{\left(2 l + 1 \right)}{l\left(l+1\right)} \left(\frac{l}{l_{0}}\right)^{n_{\sigma} - 1} }{\displaystyle\sum_{l=l_{min}}^{l_{max}} \frac{\left(2 l + 1 \right)}{l\left(l+1
\right)}  }  ,\ee
where $\xi = \hat{\cP}_{asy}/\cP_{inf}$ is negative. 
This assumes that $l(l+1)C_{l}$ for the inflaton perturbation is approximately constant over the range of $l$ 
\cite{agen1}. The large-angle asymmetry, $A_{large}$, is then given by \eq{e5} with 
$l_{min} = 2$ and $l_{max} = 64$.  On smaller angular scales, the asymmetry must be suppressed \cite{sh,notari};  for $l = 601-2048$ it must satisfy $A < 0.0045$ (95 $\%$ c.l.) \cite{sh}. For large $l$ we can replace the sums in \eq{e5} by integrals over $l$, therefore 
\be{e6} A \approx 
\frac{|\xi|_{0} (\Delta \cP_{asy}/\hat{\cP}_{asy})_{0}}{2} 
\times 
\frac{ \left(\left(\frac{l_{max}}{l_{min}}\right)^{n_{\sigma} - 1} - 1 \right)}{\left(n_{\sigma} - 1\right) \ln \left(\frac{l_{max}}{l_{min}}\right) } 
\left(\frac{l_{min}}{l_{0}}\right)^{n_{\sigma} - 1}   ~,\ee
The small-angle asymmetry, $A_{small}$,  is then given by
\eq{e6} with $l_{min} = 601$ and $l_{max} = 2048$.

\subsection{Negative Running and Asymmetries of the Spectral Index} 

The model predicts shifts of the spectral index and the running of the spectral index from their inflation model values, due to the scale-dependence of $\hat{\cP}_{asy}$.
The spectral index and its running are given by $n_{s} = n_{s\;inf} + \Delta n_{s}$ and $n_{s}^{\prime} = n_{s\;inf}^{\prime} + \Delta n_{s}^{\prime}$, 
where, to leading order in $\xi$ and neglecting the scale-dependence of $n_{s\;inf}$ \cite{agen1},  
\be{e9} \Delta n_{s} \approx \xi (n_{\sigma} - 1)  ~\ee
and
\be{e10} \Delta n_{s}^{\prime} \approx \xi (n_{\sigma} - 1)^2  ~.\ee
$\xi$ is negative, therefore the shift of the running of the spectral index is negative and the shift of the spectral index is positive if $n_{\sigma} < 1$.

 The model also predicts the existence of hemispherical asymmetries of $n_{s}$ and $n_{s}^{\prime}$ \cite{agen1}. The spectral index and its running from averaging the CMB over a hemisphere in the direction $\hat{\bfp}$ are given by 
$n_{s\;\pm} = n_{s} \pm \delta n_{s}$ and $n_{s\;\pm}^{\prime} = n_{s}^{\prime} \pm \delta n_{s}^{\prime}$, where 
$+$ ($-$) denotes the hemisphere $\theta \in $ 0 to $\pi/2$ ($\pi/2$ to $\pi$) and $\hat{\bfn}.\hat{\bfp} = \cos \theta$. $\delta n_{s}$ and $\delta n_{s}^{\prime}$ are given by \cite{agen1}
\be{e25} \delta n_{s} \approx \frac{\xi (\Delta \cP_{asy}/\hat{\cP}_{asy})}{2} \left(n_{\sigma} - 1\right) ~.\ee 
and 
\be{e27} \delta n_{s}^{\prime}  \approx \frac{\xi (\Delta \cP_{asy}/\hat{\cP}_{asy})}{2} \left(n_{\sigma} - 1\right)^{2}   
~.\ee

\begin{figure}[htbp]
\begin{center}
\epsfig{file=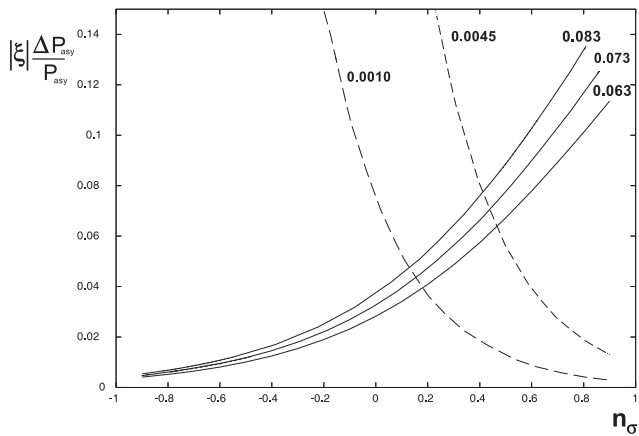, width=0.32\textwidth, angle = -90}
\caption{$|\xi| (\Delta {\cal P}_{asy}/\hat{\cP}_{asy})$ versus $n_{\sigma}$ for $A_{large} = $ 0.063, 0.073 and 0.083 (solid lines). Upper bounds on $n_{\sigma}$ from the requirement that $A_{small} < 0.0045$ and 0.0010 are also shown (dashed lines). All values are at $k = 0.002 \; {\rm Mpc}^{-1}$.  }
\label{fig1}
\end{center}
\end{figure}

\begin{figure}[htbp]
\begin{center}
\epsfig{file=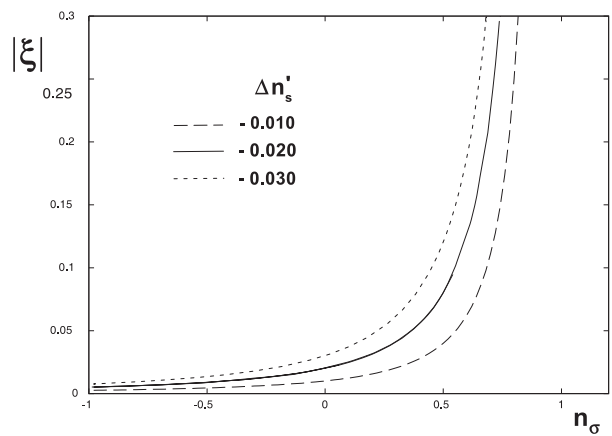, width=0.32\textwidth, angle = -90}
\caption{$|\xi|$ versus $n_{\sigma}$ for $\Delta n_{s} = $ $-0.010$, $-0.020$ and $-0.030$ when $A_{large} = 0.073$. All values are at $k = 0.002 \; {\rm Mpc^{-1}}$.}
\label{fig2}
\end{center}
\end{figure}

\begin{figure}[htbp]
\begin{center}
\epsfig{file=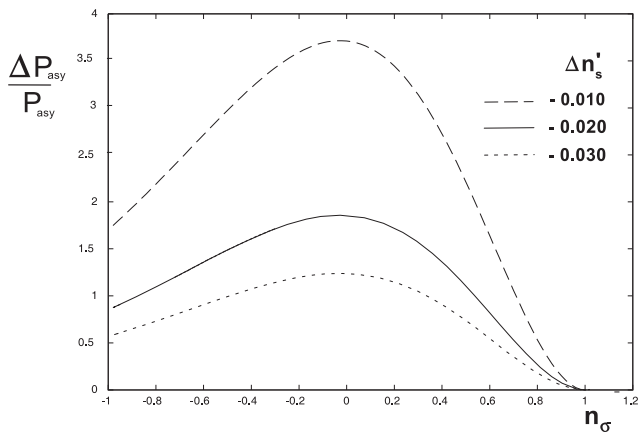, width=0.32\textwidth, angle = 90}
\caption{ $\Delta {\cal P}_{asy}/\hat{\cP}_{asy}$ versus $n_{\sigma}$ for $A_{large} = 0.073$ and $\Delta n_{s} = $ $-0.010$, $-0.020$ and $-0.030$. All values are at $k = 0.002 \;  {\rm Mpc^{-1}} $.}
\label{fig3}
\end{center}
\end{figure}

\begin{table*}
\begin{center}
\begin{tabular}{|c|c|c|c|c|c|c|}
 \hline $n_{\sigma}$	 & $\xi$ &  $\Delta n_{s}^{\prime}$ & $\Delta n_{s}$   & $(\Delta \cP_{asy}/\hat{\cP}_{asy})$ &$\delta n_{s}$ & $\delta n_{s}^{\prime}$  \\
\hline \hline	$0.44$	& $-0.032$ &	$-0.010$ & $0.018$ 	 & $2.28$ & 0.020 & -0.011  \\
\hline	$0.44$	& $-0.064$ &	$-0.020$ & $0.036$ 	 & $1.14$  & 0.020 & -0.011 \\
\hline	$0.44$	& $-0.096$ &	$-0.030$ & $0.054$ 	 & $0.76$  & 0.020 & -0.011 \\
 \hline \hline	$0.0$		& $-0.010$ & $-0.010$ & $0.010$ 	 & $3.30$  & 0.016 & -0.016 \\
\hline	$0.0$	&    $-0.020$	& $-0.020$ & $0.020$ 	 & $1.65$  & 0.016 & -0.016 \\
\hline	$0.0$	&	$-0.030$ &  $-0.030$ & $0.030$ 	& $1.10$   & 0.016 & -0.016 \\
 \hline 
\hline	$-0.5$	&	$-0.0044$ &  $-0.010$ & $0.0067$ 	& $2.72$ & 0.009 & -0.014  \\
\hline	$-0.5$	&	$-0.0088$ & $-0.020$ & $0.013$ 	&  $1.36$  & 0.009 & -0.014  \\
\hline	$-0.5$	& $-0.013$ &	$-0.030$ & $0.020$ 	&  $0.92$  & 0.009 & -0.014  \\
\hline     
 \end{tabular} 
 \caption{\footnotesize{ $\Delta n_{s}$, $\Delta n_{s}^{\prime}$ and $\Delta \cP_{asy}/\hat{\cP}_{asy}$ as a function of $n_{\sigma}$ and $\xi$ when $A_{large} = 0.073$. The corresponding hemispherical asymmetries of the spectral index and its running are also shown. 
}  }
 \end{center}
 \end{table*}

\begin{table*}
\begin{center}
\begin{tabular}{|c|c|c|c|}
 \hline $n_{\sigma}$	  & $n_{s}$ &  $\Delta n_{s}(0.05 \; {\rm Mpc}^{-1})$ &  $\Delta n_{s}^{\prime}(0.05 \; {\rm Mpc}^{-1})$  \\
\hline \hline	$0.44$ & $1.000$  & $5.9 \times 10^{-3}$ 	& $-3.3 \times 10^{-3}$  \\
\hline	$0.0$	 & $0.984$ &  $8.0 \times 10^{-4}$ & $-8.0 \times 10^{-4}$ \\
\hline	$-0.5$ & $0.977$ & $1.0 \times 10^{-4}$ & $-1.6 \times 10^{-4}$  \\
\hline     
 \end{tabular} 
 \caption{\footnotesize{Predictions for the spectral index $n_{s}$ at $k = 0.002 \; {\rm Mpc}^{-1}$ as a function of $n_{\sigma}$ when $n_{s}^{\prime} = -0.020$ for the $\phi^2$ chaotic inflation model. The shifts of the spectral index and its running at $k = 0.05 \; {\rm Mpc}^{-1}$ are also shown.     
}  }
 \end{center}
 \end{table*}

\section{Results and Conclusions}      

   The model has three input parameters, which can be taken to be $n_{\sigma}$, $\xi$ and $\Delta \cP_{asy}/\hat{\cP}_{asy}$. These can determine five observable quantities: $A$, $\Delta n_{s}$, $\Delta n_{s}^{\prime}$, $\delta n_{s}$ and $\delta n_{s}^{\prime}$. Therefore it is always possible to predict two observables by    
using three as inputs. This will allow us to test the consistency of the power spectrum model. 
$A$, $\delta n_{s}$ and  $\delta n_{s}^{\prime}$ are in fact  determined by just two parameters, $n_{\sigma}$ and $\xi (\Delta \cP_{asy}/\hat{\cP}_{asy})$, therefore to fix all three input parameters one of either $\Delta n_{s}$ or $\Delta n_{s}^{\prime}$ is also necessary. 
In the following all scale-dependent quantities are given by their values at $k = 0.002 \; {\rm Mpc}^{-1}$ unless otherwise stated. This is chosen to represent low multipoles ($l = 28$) where the effects of $\cP_{asy}$ are largest. 

    We first illustrate the parameter space for which the model can account for the asymmetry and the negative running of the spectral index. In Figure 1 we show $|\xi| (\Delta {\cal P}_{asy}/\hat{P}_{asy})$ as a function of $n_{\sigma}$ for a range of $A_{large}$ corresponding to the 1-$\sigma$ limits\footnote{$|\xi| (\Delta {\cal P}_{asy}/\hat{P}_{asy})$ is equivalent to $|\Delta \cP_{asy}|/\cP_{inf}$.} . We also show the upper limit on $n_{\sigma}$ from the requirement that $A_{small} < 0.0045$. For $A_{large} = 0.073$ this requires that $n_{\sigma} < 0.44$. To show the effect of a tighter upper bound on $A_{small}$, we also show the upper limit on $n_{\sigma}$ for the case $A_{small} < 0.0010$.  In Figure 2 we show $|\xi|$ as a function of $n_{\sigma}$ for a range of $\Delta n_{s}^{\prime}$, corresponding to the Planck 1-$\sigma$ limits in the case with tensor modes. In Figure 3 we show the corresponding values of $\Delta {\cal P}_{asy}/\hat{\cP}_{asy}$.

  We next show, by explicit example, that it is possible to choose values for the model parameters which account for the observed hemispherical asymmetry, with the necessary suppression at large multipoles, while generating a running of the spectral index of the right magnitude to account for the Planck observation and also to allow the Planck upper bound on $r$ to be consistent with a value significantly larger than 0.1. We also illustrate how the model might be tested.

     In Table 1 we give $\Delta n_{s}$, $\Delta n_{s}^{\prime}$ and $\Delta \cP_{asy}/\hat{\cP}_{asy}$ as a function of $n_{\sigma}$ and $\xi$ when $A_{large} = 0.073$.
The shifts $\Delta n_{s}^{\prime}$ in Table 1 are large enough ($\Delta n_{s}^{\prime} \sim -0.020$) to bring the Planck upper bound on $r$ into agreement with a value significantly larger than 0.1. 
We also give the corresponding asymmetries $\delta n_{s}$ and $\delta n_{s}^{\prime}$, which depend only on $n_{\sigma}$ when $A_{large}$ is fixed.

In order to predict the observed spectral index,
$\Delta n_{s}$ must be added to values for $n_{s\;inf}$ and $n_{s\;inf}^{\prime}$ from a specific inflation model.  As an example we will consider $\phi^2$ chaotic inflation. For this model, at a given scale $k$, 
$n_{s\;inf} = 1- 2/N_{*} = 0.964$, $n_{s\;inf}^{\prime} = -2/N_{*}^{2} = - 0.00066$ and $r = 8/N_{*} = 0.15$, where $N_{*}$ is the number of e-foldings at which the scale $k$ exits the horizon and we have assumed $N_{*} = 55$ in our numerical values. In Table 2 we give the total spectral index $n_{s}$ at $k = 0.002 \; {\rm Mpc}^{-1}$ for the $\phi^2$ model as a function of $n_{\sigma}$ when $n_{s}^{\prime} = -0.020$ and $A_{large} = 0.073$. The value of $n_{s}^{\prime}$ ($\equiv \Delta n_{s}^{\prime}$ in the $\phi^2$ inflation  model) is chosen to be of the order of magnitude required by Planck for consistency with a value of $r$ significantly larger than 0.1. Therefore this model can account for the hemispherical asymmetry, the negative running of the spectral index and consistency with a large value of $r$. The value of $n_{\sigma}$ can be fixed by observing the asymmetry of the spectral index $\delta n_{s}$, since from \eq{e5} and \eq{e25} $n_{\sigma}$ can be fixed by $A_{large}$ and $\delta n_{s}$. We also give the shift of the spectral index and its running at the Planck pivot scale $k_{0} = 0.05 \; {\rm Mpc}^{-1}$. Since this corresponds to a high multipole number, $l_{0} \approx 700$, $\cP_{asy}$ is strongly suppressed and so the shifts are small. Therefore the spectral index at the Planck pivot scale is essentially $n_{s\;inf}$. For the $\phi^2$ model this is in good agreement with the 1-$\sigma$ Planck observation,  $n_{s} = 0.9607 \pm 0.007$ \cite{cosmoparams}.

  In conclusion, a modulation of the adiabatic power spectrum from inflation which suppresses the power at low multipoles can account both for the hemispherical asymmetry of the CMB power spectrum and for a negative running of the spectral index of the magnitude observed by Planck. The latter can also allow the Planck upper bound on $r$ to be consistent values significantly larger than 0.1, as originally reported by BICEP2. The model is predictive and in principle testable.

Our analysis is based on a modulated power spectrum for the adiabatic perturbation. The next step will be to find a realization of such a power spectrum in an explicit field-based model. A strong constraint on such models is the observed isotropy of the CMB mean temperature. This implies that the spatial modulation of the inflaton power spectrum must be due to modulation of the fluctuations of the energy density, leaving the mean energy density unaffected. This can be achieved by a space-dependent sound speed for the inflaton due, for example,  to a spatially-varying Dirac-Born-Infeld kinetic term \cite{cai}.

 \section*{Acknowledgements}
The work of JM is partially supported by STFC grant ST/J000418/1.

\end{document}